\documentclass{article}
\usepackage{graphicx,amssymb,amsmath,geometry,amsthm}
\newtheorem{theorem}{Theorem}
\newtheorem{lemma}{Lemma}
\newtheorem{obs}{Observation}

\title{Improved Algorithm for Computing the Maximum-volume Bichromatic Separating Box}

\author{
Bogdan Armaselu \\
{\tt barmaselub@gmail.com}
}
\date{}

\index{Armaselu, Bogdan}

\begin{document}
\thispagestyle{empty}
\maketitle

\begin{abstract}
We consider the problem of computing the largest-area bichromatic separating box among a set of $n$ red points and a set of$m$ blue points in three dimensions.
Currently, the best known algorithm to solve this problem takes $O(m^2 (m + n))$ time and $O(m + n)$ space.
In this paper, we come up with an improved algorithm for the problem, which takes $O(m^2 + n)$ time.
\end{abstract}

\section{Introduction}
\label{s:intro} 

Consider a set $R$ consisting of $n$ set points and a set $B$ consisting of $m$ blue points in the three dimensional space.
A \textit{separating box} for $R$ and $B$ is a box (hyper-rectangle) that contains all the red points in $R$ and the fewest blue points in $B$.
We call a separating box a \textit{maximum-volume bichromatic separating box} (MBSB) of $R$ and $B$, or simply \textit{maximum separating box}, if it has the largest volume among all boxes separating $R$ and $B$.
The problem was introduced in \cite{Armaselu-CCCG}.
In this paper, we improve the running time of $O(m^2 (m + n))$ in \cite{Armaselu-CCCG} and provide an algorithm that takes $O(m^2 + n)$ to find one optimal solution.

\begin{figure}[t]
\centering
\includegraphics[scale=0.33]{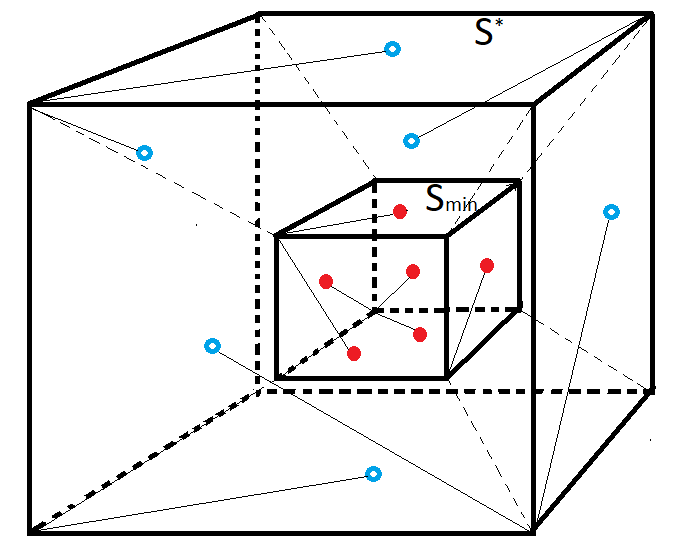}
\caption{Red points in $R$ are shown in spheres and blue points in $B$ are shown in donuts. 
The minimum enclosing box $S_{min}$ of all red points, as well as the maximim volume separating box $S^*$ of $R$ and $B$, are shown. 
All thin solid straight lines are embedded in a face of a box and all thin dotted straight lines connect corresponding corners of $S^*$ and $S_{min}$.}
\label{fig:example}
\end{figure}

An example of MBSB in shown in Figure \ref{fig:example}.

There are quite a few interesting real world applications of this problem.
For instance, in oncology, one may be interested in separating a tumor from the healthy three-dimensional tissue.
The tumor cells are indicated by red points, while the healthy cells and background stoma are indicated by blue points.
Often times, an oncology pathologist may seek to surgically cut out the largest tissue containing the tumor while minimizing healthy tissue removall, which leads to the problem we solve.

Another application would be in spatial databases.
Suppose the goal is to retrieve all "red" data from a table with 3 real-valued columns and a discrete label column, which can be "red" or "blue", while minimizing the number of records with "blue" label retrieved.
Similarly, there are applications in data science and machine learning, where we are given a dataset with 3 real-valued features and two classes, red and blue, and the goal is to train a classifier that can classify the data as red or blue.
If much higher weight is assigned to red misclassifications than to blue misclassifications, then computing the largest separating box of the point sets is important, as it misses no red points.

\subsection{Related Work}
\label{ss:related-work} 

Geometric separability is a very important problem in the field of geometric optimization.
Strong separability problems ask for perfect separators, where no misclassifications are allowed, whereas in weak separability problems, the goal is to minimize an error measure related to classification.
Error measures may include number of misclassifications, total misclassification weight, or other.
For instance, computing the largest (maximum-volume) separating box is a weak separability problem which minimized the number of misclassified blue points.

Perhaps the most commonly studied separability topic is the linear separability, in which the separator is a line, a plane, or a hyperplane.
The hyperplane separability problem was introduced by Megiddo et. al \cite{Megiddo}.
Given a set of red points and a set of blue points, the goal is to find a hyperplane such that points from no more than one set can be found on either side of the hyperplane.
They show how to solve the decision version of the problem in linear time using linear programming.
In addition, they prove that $k$-line separability, which ask whether two points sets are separable by $k$ lines, is NP-complete.
Aronov et. al \cite{Aronov} studied linear separability of $n$ points in $d \geq 2$ dimensions using 4 error metrics, one of which is ther total number of misclassified points.
The other metrics measure the maximum euclidean distance of a misclassified point to the hyperplane, the sum of these distances, and the squared sum of these distances, respectively.
For minimizing the number of misclassifications, they give an $O(n^d)$ time algorithm.
Later on, Chan \cite{Chan} solved the relaxed version of the problem in 2 and 3 dimensions, in which up to $k$ violations are allowed.
Their algorithm for the 2D version runs in $O((n + k^2) \log n)$ time, while their 3D algorithm runs in $O(n + k^{\frac{11}{4}} n^{\frac{1}{4}})$.
Dobkin solved the problem of linear separability of polyhedra, for which they provide a linear time algorithm \cite{Dobkin}.

Circular separability was introduced by Fisk \cite{Fisk}.
They study the optimization version, in which the goal is to find the separating circle with no violations and of minimum radius.
They provide a $O(n^2)$ time algorithm involving nearest and farthest Voronoi diagrams.
Later on, O'Rourke et. al improved the result to linear time \cite{O'Rourke} for the decision problem, and $O(n \log n)$ time for the optimization problem.
In the weak circular separability problem, the minimum-radius circle containing all of the $n$ red points and the fewest of the $m$ blue points is sought.
Weak circular separability was introduced by Bitner et. al \cite{Bitner}, who presented two solutions.
One of them is based on furthest-point Voronoi Diagrams and requires $O(m n \log m + n \log n)$ time and $O(m + n)$ space.
The other involves circular range queries and takes $O((m + n) \log n + m^{1.5} \log^{O(1)}m)$ time and $O(m^{1.5} \log^{O(1)}m)$ space.
Later on, Armaselu et. al \cite{Armaselu-TCS} solved the dynamic version of the problem, in which blue points are inserted and removed dynamically.
They provide three data structures, one of which requires $O(m + n)$ space and can answer insertion queries in $O(n + \log m)$ time, deletion queries in $O(n \log m)$ time, and is updated in $O((m + n) \log m)$ time.
The other two are designed for insertions only (respectively, deletions only) and can answer queries in $O(\log (m n))$ time, at the expense of an $O(m n)$ space requirement and $O(m n \log^{O(1)} (m n))$ time requirement.

Rectangular separability was studied by Armaselu and Daescu.
They provide a $O(m \log m + n)$-time algorithm for the 2D axis-aligned version of the problem, which asks for the largest area axis-aligned rectangle separating red and blue points in the plane \cite{Armaselu-arXiv}.
They also solve the planar arbitrary orientation version, in which the target rectangle has arbitrary orientation, in $O(m^3 + n \log n)$ time, and the 3D axis-aligned version in $O(m^2 (m + n))$ time \cite{Armaselu-CCCG}, which we improve in this paper.

An interesting related problem is the problem of computing the largest empty rectangle among a set of $n$ planar points, introduced by Hsu et. al \cite{Hsu}.
They solve trhe axis-aligned version with an algorithm that takes $O(n^2)$ worst-case, $O(n \log^2 n)$ expected time to find all optimal solutions.
Later, Chazelle et. al improved the running time to $O(n \log^3)$ time to find one optimal solution.
Aggarwal et. al \cite{Aggarwal} further improved the running time to $O(n \log^2 n)$ to find one optimal solution, which is currently still the best known algorithm for the axis-aligned version.
For the arbitrary orientation version, Mukhopadhyay et. al \cite{Mukhopadhyay} and Chaudhuri et. al \cite{Chaudhuri} independenlty provided algorithms to find all optimal solutions in $O(n^3)$ time and $O(n^2)$ space.
The 3D axis-aligned case was introduced by Nandy et. al, who give a cubic time, linear space algorithm to find all optimal solutions \cite{Nandy}.
Kaplan et. al considered the problem of computing the largest empty rectangle containing query point \cite{Sharir}.
They come up with a data structure that answers queries in $O(log^4 n)$ time, with $O(n \alpha(n) \log^4 n)$ time preprocessing, where $\alpha(n)$ is the very slow-growing inverse Ackermann function.

Various other separators have been considered by Hurtado et. al, such as wedges, strips \cite{Mora}, and double wedges \cite{Hurtado} in 2D.
All these separation problems can be decided in $O(n \log n)$ time \cite{Hurtado, Mora}.
They also extend the single and double wedge separability to 3D and consider separability by other 3D geometric loci, such as prisms and pyramids \cite{Seara}, for which they describe polynomial-time algorithms.
Demaine et. al studied the separability by line segments, chords, and multi-chords \cite{Demaine}, for which they give polynomial time algorithms and also prove lower bounds.

\subsection{Our Results}
\label{ss:results}

In this paper, we improve the running time of the algorithm in \cite{Armaselu-CCCG} for computing an MBSB, 
using a more clever approach which breaks down all possible configurations of optimal solutions into 6 cases, and uses clever tricks to solve the most complex of these cases in $O(m^2)$ time, after the smallest $R$-enclosing rectangle $S_{min}$ is computed.
In total, our algorithm takes $O(m^2 + n)$ time, which is an improvement of at least a factor of $m$ compared to the previous known bounds.

The rest of the paper is structured as follows.
In Section \ref{s:prelim}, we make some importnant preliminary observations.
Then, in section \ref{s:cases}, we analyze all the situations in which a largets separating box can be found, provide a detailed description of the algorithm, and analyze its running time.
Finally, in Section \ref{s:conclusion}, we draw the conclusions and list some future directions.

\section{Preliminaries}
\label{s:prelim}

\begin{figure}[t]
\centering
\includegraphics[scale=0.45]{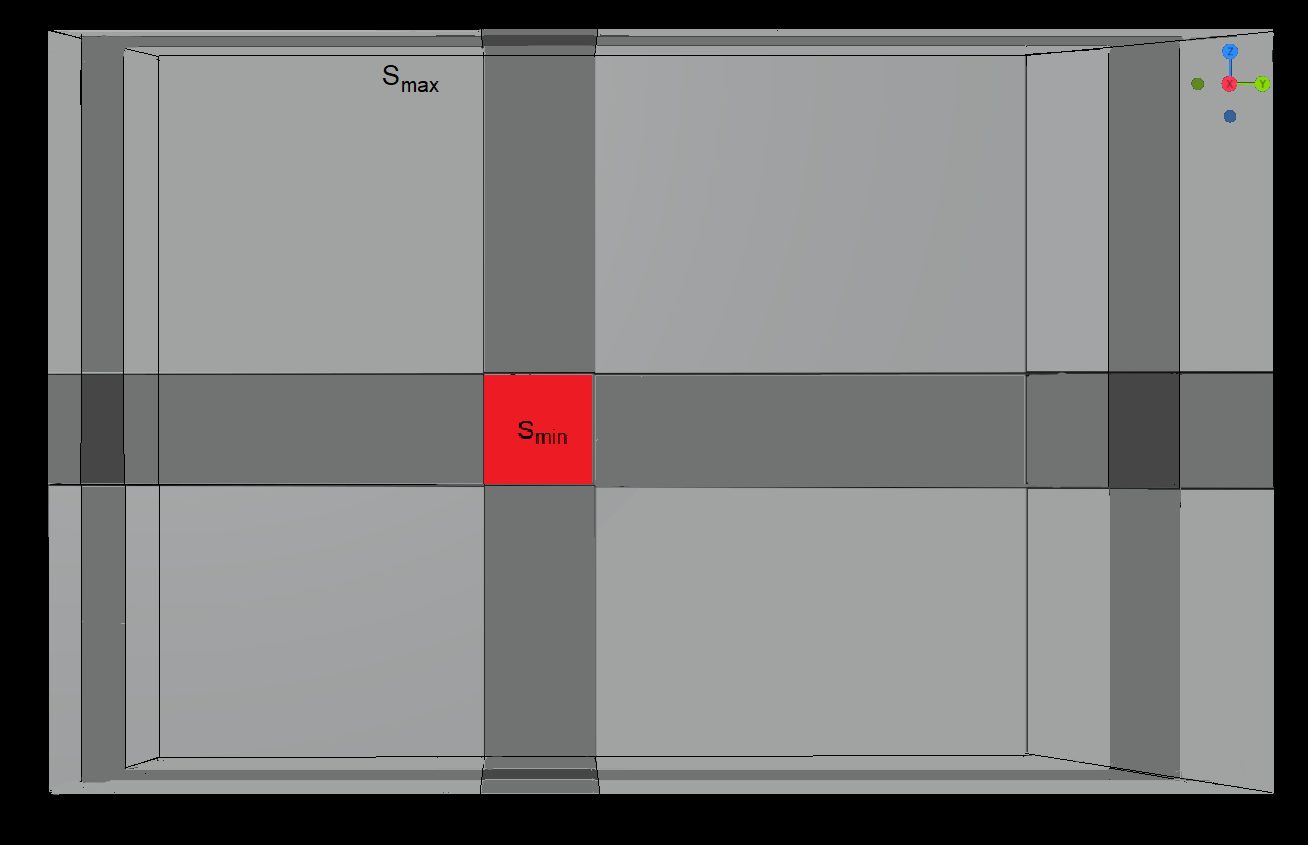}
\caption{The boxes $S_{min}$ (shown in red) and $S_{max}$, along with the planes defining them, partition the space into 8 corner regions (shown in light gray), 6 side regions, and 12 edge regions (shown in dark gray).
Note that only the front side of $S_{min}$ is shown, and the plane defining the front face of $S_{min}$ occludes everything in its rear side (including all the side regions).
Notice the axis legend at the top right corner of the image.}
\label{fig:strip}
\end{figure}

We begin by computing the smallest $R$-enclosing box $S_{min}$ and disarding the blue points inside $S_{min}$, and they are inevitable.
Then we extend $S_{min}$ outwards in each direction parallel to itself until it hits a blue point, and denote the resulting box by $S_{max}$, as in \cite{Armaselu-CCCG}.
The planes that define $S_{min}$ partition $interior(S_{max})$ into $S_{min}$ and 26 other regions (see Figure \ref{fig:strip} for an illustration).
Specifically, there are:

\begin{itemize}
\item 8 \textit{corner} regions, which are boxes cornered by one corner of $S_{min}$ and the corresponding corner of $S_{max}$ 
(e.g. top-right-front (TRF) corner of $S_{min}$ and TRF corner of $S_{max}$, bottom-left-back (BLK) corner of $S_{min}$ and BLK corner of $S_{max}$)

\item 6 \textit{side} regions, which are boxes bounded by one face of $S_{min}$ (e.g. top face) and the corresponding face of $S_{max}$ (e.g. top face)

\item 12 \textit{edge} regions, which are boxes bounded by one edge of $S_{min}$ (e.g. top-right or TR) and the corresponding edge of $S_{max}$.
\end{itemize}

Note that the side regions may not contain any blue points, otherwise they would redefine $S_{max}$.
We sort all blue points first by $X$, then by $Y$, and then by $Z$ coordinate.

A \textit{candidate box} is a separating box that cannot be extended in any direction without including a blue point.
Every candidate box is supported by 6 blue points, called \textit{supports}, each of them supporting a face of the box.

A \textit{halfspace of $S_{min}$} is a halfspace $H$ bounded by the plane $L$ supporting a face $F$ of $S_{min}$ and not containing $S_{min}$.
A halfspace $H$ of $S_{min}$ is said to be \textit{perpendicular} to another halfspace $H^{\perp}$ of $S_{min}$, and we write $H \perp H^{\perp}$,
if $H^{\perp}$ is bounded by a plane $L^{\perp} \perp L$ supporting a face $G$ of $S_{min}$ and not containing $S_{min}$.

If $H$ is a halfspace of $S_{min}$, we denote the face of $S_{min}$ supporting the plane bounding $H$ by $F_H$, and the complement of $H$ by $C_H$.

We call a combination of 2 supports a \textit{pair}, a combination of 3 supports a \textit{triple}, and a combination of 4 supports a \textit{quadruple}.

Let $RP$ be the maximal rectilinear polyhedron that contains all red points and none of the blue points, i.e. $RP$ cannot be extended in any direction without including a blue point.

By convention, we consider a point $p$ to be \textit{in front of} (resp., \textit{to the back of}) another point $q$ if $y(p) > y(q)$ (resp., $y(p) < y(q)$), where $y(r)$ is the Y coordinate of point $r$.
We also consider that the scene is viewed by a camera located at $Y = -\infty$.
That is, the \textit{near} direction is towards -Y while the \textit{far} direction is towards +Y.
Similarly, $p$ is considered \textit{above} (resp., \textit{below}) another point $q$ if $z(p) > z(q)$ (resp., $z(p) < z(q)$).
Finally, $p$ is considered \textit{to the right of} (resp., \textit{to the left of}) another point $q$ if $x(p) > x(q)$ (resp., $x(p) < x(q)$).

\begin{figure}[t]
\centering
\includegraphics[scale=0.55]{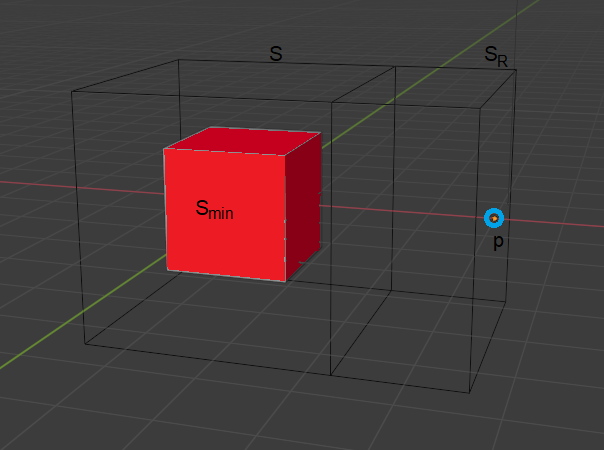}
\caption{Box $S$ is extended to the right until it hits a blue point $p$ on its right face, resulting in a new box $S_R$.}
\label{fig:right-support}
\end{figure}

\begin{obs}
\label{obs:prelim-1}
For every face $F$ of a candidate box $S$, once adjacent supports are established, the $F$ support can be found by extending $S$ in the direction of $F$ until it hits a blue point.
For instance, once the top and back supports of $S$ are established, the right support of $S$ can be found by extending $S$ to the right until it hits a blue point (see Figure \ref{fig:right-support} for details).
\end{obs}

\begin{figure}[t]
\centering
\includegraphics[scale=0.45]{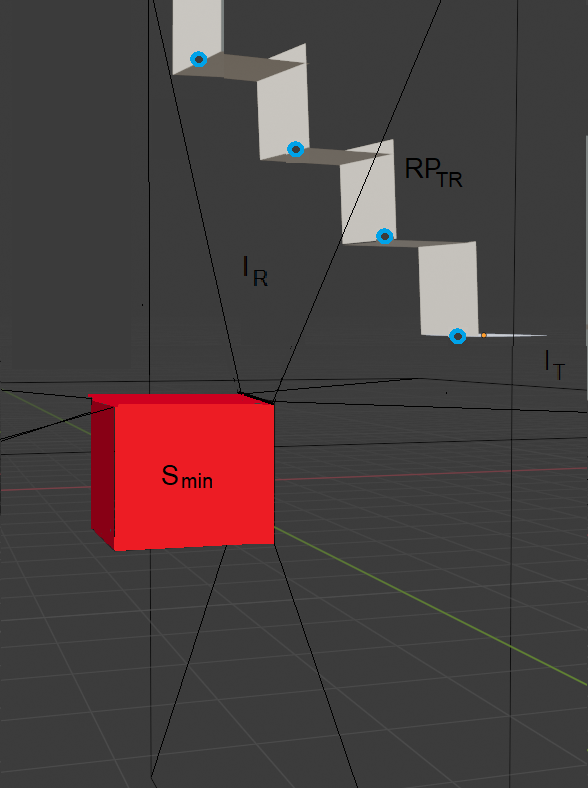}
\caption{Planes $l_R$ and $l_T$ bounding the right (resp., top) faces of $S_{min}$ define the top-right edge region, which contains 4 blue points. 
The chain of plains denotes $RP_{TR}$, the portion of $RP$ within the top-right edge region.
Notice that each blue point defines an edge of $RP_{TR}$, together forming a staircase.}
\label{fig:edge-staircase}
\end{figure}

\begin{obs}
\label{obs:prelim-2}
For each edge region $E$, the portion of $RP$ within $E$, denoted $RP_E$, has the shape of a prism with a staircase-shaped base (as shown in Figure \ref{fig:edge-staircase}).
\end{obs}

\begin{figure}[t]
\centering
\includegraphics[scale=0.45]{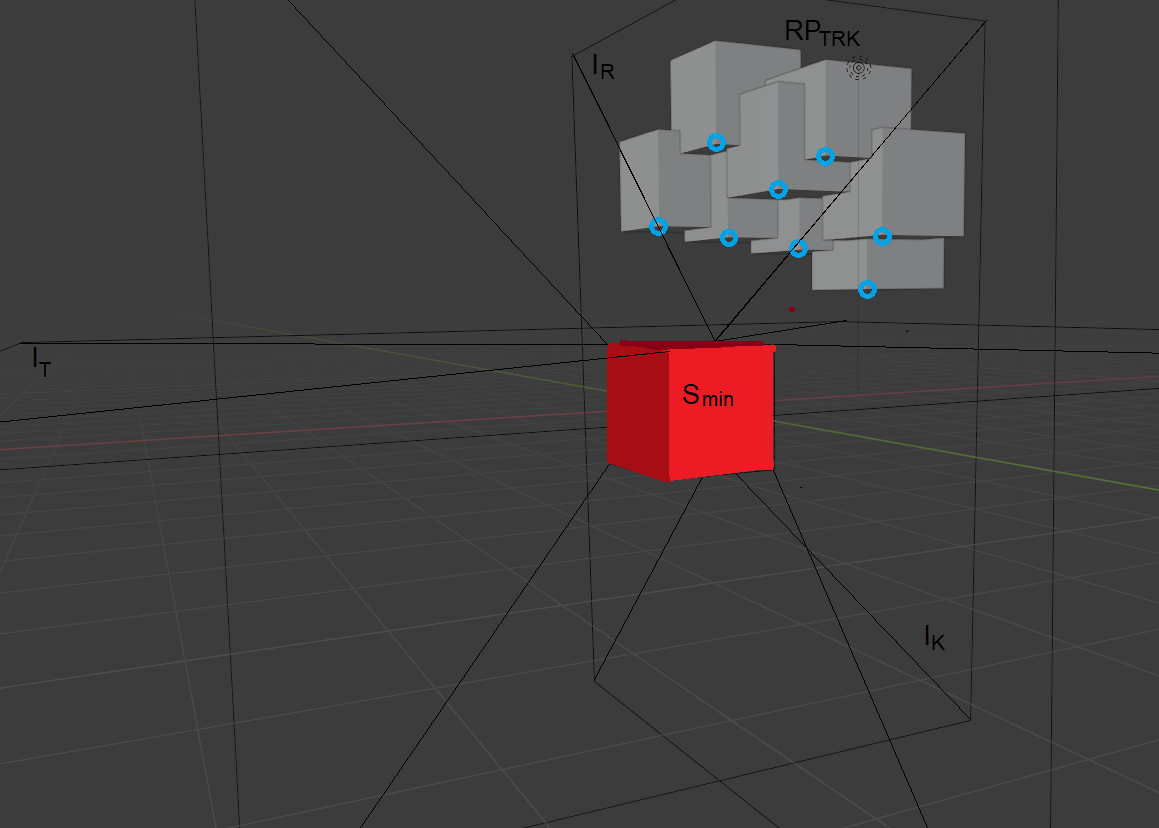}
\caption{Planes $l_R, l_T, l_K$ bounding the right (resp., top, back) faces of $S_{min}$ define the top-right-back corner region, which contains 8 blue points. 
The rectilinear polyhedron denotes $RP_{TRK}$, the portion of $RP$ within the top-right-back corner region.
Notice that each blue point defines an corner of $RP_{TRK}$.
Similarly, each blue point from the bottom-left-front region defines a point of $RP_{BLF}$ (not shown in image)}
\label{fig:corner-polyhedron}
\end{figure}

\begin{obs}
\label{obs:prelim-3}
For each corner region $C$, the portion of $RP$ within $C$, denoted $RP_C$, has a three-dimensional staircase shape (as displayed in Figure \ref{fig:corner-polyhedron}).
\end{obs}

\begin{obs}
\label{obs:prelim-4}
For every corner region $C$, there are $O(m)$ triples with all 3 supports from $C$.
\end{obs}

From now on, for each corner region $C$, whenever understood, we will refer to $RP_C$ as the \textit{polyhedron of $C$}.
In addition, for each edge region $E$, we will refer to $RP_E$ as the \textit{staircase of $E$}.

\section{A breakdown of the situations a candidate box can be found in}
\label{s:cases}

Note that:
\begin{itemize}
\item no more than 5 supports of a candidate box can be located in any halfspace of $S_{min}$,

\item no more than 4 supports can be located in the intersection of two halfspaces of $S_{min}$,

\item no more than 3 supports can be located in two staircases within a halfspace of $S_{min}$, and

\item no more than 3 supports can be found in any corner region.
\end{itemize}

Based on the location of the supports, a candidate box $S$ can be in one of the following cases.

\begin{itemize}
\item \textbf{Case 1}. 5 supports in a halfspace $H$ of $S_{min}$, out of which 4 in a halfspace $H^{\perp} \perp H$ of $S_{min}$.

\item \textbf{Case 2}. 5 supports in a halfspace $H$ of $S_{min}$, out of which 3 in a halfspace $H^{\perp} \perp H$ of $S_{min}$.

\item \textbf{Case 3}. 5 supports in a halfspace $H$ of $S_{min}$, out of which 2 in a halfspace $H^{\perp} \perp H$ of $S_{min}$, and at least one in an edge region $E \subset H - H^{\perp}$ adjacent to a corner region within $H^{\perp}$

\item \textbf{Case 4}. 5 supports in a halfspace $H$ of $S_{min}$, out of which 3 in two edge regions $E_1, E_2 \subset H - H^{\perp}$ adjacent to a corner region within $H^{\perp}$

\item \textbf{Case 5}. 4 supports in a halfspace $H$ of $S_{min}$, 3 of which in the complement $C_{H^{\perp}}$ of a halfspace $H^{\perp} \perp H$ of $S_{min}$.

\item \textbf{Case 6}. 4 supports in a halfspace $H$ of $S_{min}$, with one support in each corner region within $H \cap H^{\perp}$.

\item \textbf{Case 7}. 4 supports in a halfspace $H$ of $S_{min}$, with two supports in one of the corner regions of $H \cap H^{\perp}$.

\item \textbf{Case 8}. 3 supports in 2 opposite corner regions.
\end{itemize}

In the following subsections we will analyze the number of candidate boxes in each case.
For cases 1 through 7 assume, without loss of generality, that $F_H$ is the right face of $S_{min}$ and $F_{H^{\perp}}$ is the back face of $S_{min}$.

\subsection{Case 1}
\label{ss:cases-1}

\begin{figure}[t]
\centering
\includegraphics[scale=0.55]{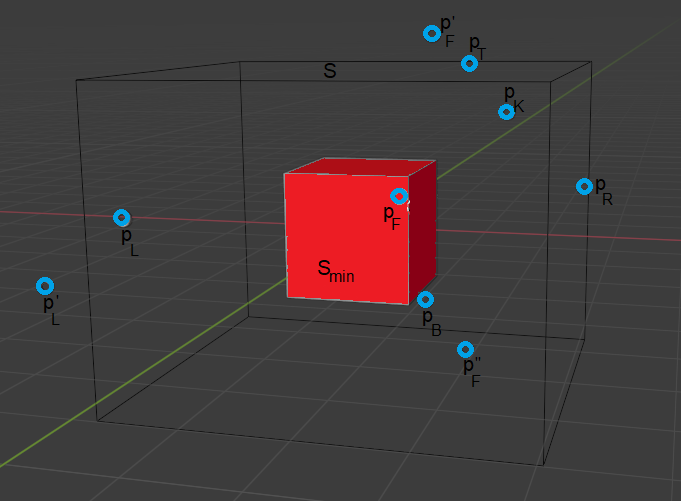}
\caption{Points $p_R, p_B, p_T, p_K$ support, respectively, the right, bottom, top, and back faces of $S$. 
The front support $p_F$, and thus the left support $p_L$, are uniquely determined, since extending $S$ to touch $p_F', p_F''$ introduces a blue point.
}
\label{fig:case1}
\end{figure}

Refer to Figure \ref{fig:case1} for an illustration of this case.

Suppose wlog that the 4 supports in $H^{\perp}$ are the top, back, right, and bottom supports for $S$.
Thus, $S$ may only have one valid front support.
Once the top, back, right, and bottom supports are determined, the front support of $S$ can be found by extending $S$ frontwards until it hits a blue point.
The left support of $S$ is now uniquely determined.
To bound the number of top-back-right-bottom quadruples coming from the back side of $F$, we note the following.
The top support cannot come from the top-back edge region, since it is outside $H^{\perp}$, so it must come from the top-back-right corner region, which makes the bottom support uniquely determined.
Since there are $O(m)$ choices of top supports, it follows that there are $O(m)$ combinations of top-back-right-bottom quadruples, which in turn implies there are $O(m)$ candidate boxes in case 1.

\subsection{Case 2}
\label{ss:cases-2}

\begin{figure}[t]
\centering
\includegraphics[scale=0.55]{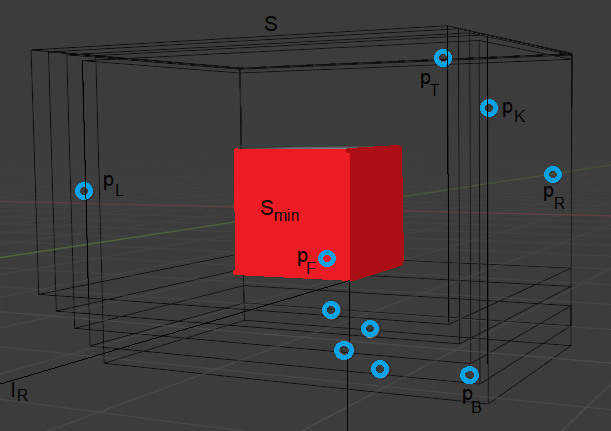}
\caption{Points $p_R, p_T, p_K$ support, respectively, the right, top, and back faces of $S$. 
There is one front-bottom pair with $p_F$ as front support, one with $p_B$ as bottom support, and $O(m)$ with front and bottom supports from $RP_{BRF}$.
}
\label{fig:case2}
\end{figure}

See Figure \ref{fig:case2} for a depiction of this case.

First, assume the 3 supports in $H^{\perp}$ to be the top, back, and bottom supports for $S$.
In this case there are $O(m)$ valid front-right pairs.
Specifically, there is one pair with the right support from the right-top or the right-bottom edge region, 
one pair with the front support from the front-top or the front-bottom edge region,
$O(m)$ pairs from the front-right staircase,
and $O(m)$ pairs from the front-right-top and front-right-bottom rectilinear polyhedra.
For each front-right pair, there is a unique left support.
To bound the number of top-back-bottom triples, note that, once the top, right, and back supports are established, the bottom support is unique.
Since the top-back pairs come from the top-right-back polyhedron, there are $O(m)$ such pairs, and for each of them we have $O(m)$ valid right-bottom-front-left quadruples.
It follows that there are $O(m^2)$ candidate boxes with top-back-bottom triples from $H^{\perp}$.

Now assume the 3 suports in $H^{\perp}$ to be the top, back, and right supports for $S$.
The situation where the supports are bottom, back, and right supports for $S$ is symmetric.
In this case there are $O(m)$ valid front-bottom pairs.
Specifically, there is one pair with the bottom support from the bottom-right or bottom-left edge region, 
one pair with the front support from the front-left or front-right edge region,
and $O(m)$ pairs from the front-right-bottom rectilinear polyhedron.
For each front-bottom pair, there is a unique left support.
As for the number of top-back-right supports, note that, once the top and back supports are established, the right support is unique.
Since the top-back pairs come from the top-right-back polyhedron, there are $O(m)$ such pairs, and for each of them we have $O(m)$ valid right-bottom-front-left quadruples.
This implies an $O(m^2)$ total candidate boxes with top-right-back triples from $H^{\perp}$. 

We conclude that $O(m^2)$ candidate boxes are in case 2.

\subsection{Case 3}
\label{ss:cases-3}

\begin{figure}[t]
\centering
\includegraphics[scale=0.55]{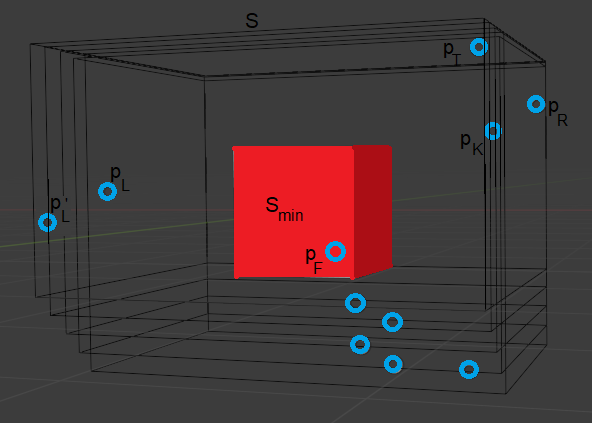}
\caption{Points $p_R, p_K \in RP_{TRK}$ support, respectively, the right and back faces of $S$. 
The top support $p_T$ is uniquely determined, which leaves $O(m)$ valid bomtom-front pairs, each of them having a unique left support.
}
\label{fig:case3}
\end{figure}

Figure \ref{fig:case3} supports the description of this case.

First, suppose wlog that the 2 supports in $H^{\perp}$ are the top and back supports of $S$.
If they come from the same polyhedron, then the right support is unique.
Since only the top and the back support are in $H^{\perp}$, the right support must come either from $E$ or from the half-space $-H^\perp$ of $S$ opposite to $H^{\perp}$.
If it comes from $E$, then the bottom support is unique, which in turn uniquely determins the front support, which finally uniquely determins the left support.
Otherwise, the bottom support must come from a staircase, which means the front and right supports must be located in $-H^\perp$.
This gives us $O(m)$ valid front-right pairs for the subcase where the top and back supports come from the same polyhedron.
If the back and top supports come from different polyhedra, then the right and the bottom supports are unique.
This in turn causes the front and the left supports to be unique.
Thus, there are $O(m)$ candidare boxes in the subcase where the 2 supports in $H^{\perp}$ are the top-back pair.

Now wlog suppose that the 2 supports in $H^{\perp}$ are the back and right supports of $S$.
If they come from a staircase, then the top and the bottom supports are unique, which in turn makes the front and the left supports unique.
Otherwise, suppose that the back support comes from a polyhedron.
If the right support comes from the same polyhedron, then the pair uniquely determines either the top or the bottom support.
As a consequence, there are $O(m)$ valid bottom-front (resp., bottom-back) pairs, and for each of them there is a unique left support.
If, on the contrary, the right support comes from a staircase or another polyhedron, then both the top and the bottom supports are established, which uniquely determines the front and the left supports.
Hence, there are $O(m)$ candidate boxes where the 2 supports in $H^{\perp}$ are the right-back pair.

Since there are $O(m)$ top-back pairs, as well as $O(m)$ right-back pairs, it follows that there are $O(m^2)$ candidate boxes in case 3.

\subsection{Case 4}
\label{ss:cases-4}

\begin{figure}[t]
\centering
\includegraphics[scale=0.55]{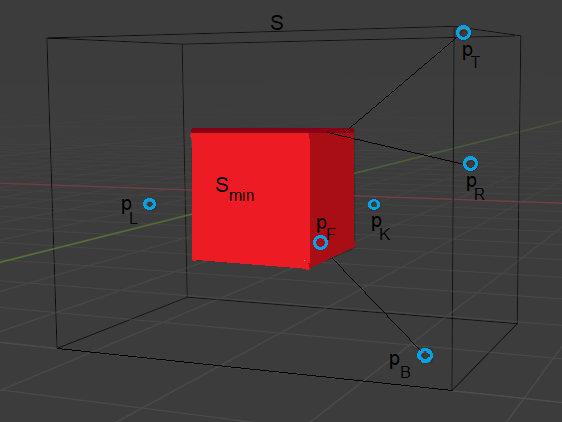}
\caption{Points $p_T, p_R \in RP_{TR}, p_B \in RP_{BR}$ support, respectively, the top, right, and bottom faces of $S$. 
Dark lines shooting from these points to $S_{min}$ have constant $Y$ coordinate.
The back support $p_K$ and the front support $p_F$ are uniquely determined, which makes the left support $p_L$ unique.
}
\label{fig:case4}
\end{figure}

Assume wlog that the top-right pair comes from the top-right edge region $E_1$ and the bottom support comes from the bottom-right edge region $E_2$.
Then the front and back supports are uniquely determined, which in turn implies that the left support is uniquely determined (see Figure \ref{fig:case4}).
Since there are $O(m)$ top-right pairs, it follows that there are $O(m)$ candidate boxes in case 4.

\subsection{Case 5}
\label{ss:cases-5}

\begin{figure}[t]
\centering
\includegraphics[scale=0.45]{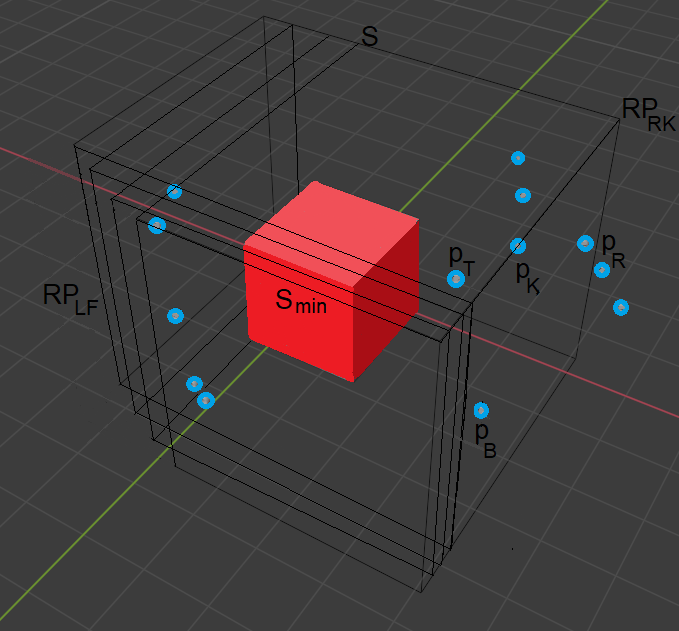}
\caption{Points $p_K, p_R \in RP_{RK}, p_T \in H^{\perp}$ support, respectively, the back, right, and top faces of $S$. 
The bottom support $p_B$ is uniquely determined.
There are $O(m)$ such right-back pairs, and for each of them there are $O(m)$ choices of left-front pairs $(p_L, p_F), p_L, p_F \in RP_{LF}$.
}
\label{fig:case5}
\end{figure}

Assume wlog that $F_H$ is the right face of $S_{min}$, and $F_{C_{H^{\perp}}}$ is the back face of $S_{min}$ (refer to Figure \ref{fig:case5}).

If two of the supports in $C_{H^{\perp}}$ are from the right-back staircase, then either the top or the bottom support is in $H^{\perp}$.
Wlog suppose the top support is in $H^{\perp}$.
If the bottom support is in $H$, then it is the only valid bottom support.
Otherwise, the front support is in $H$, and similarly it follows that it is the only valid front support.
Either way, there are $O(m)$ valid pairs involving the left support: left-front or left-bottom pairs.
Also, there are $O(m)$ right-back pairs from the right-back staircase, which gives a total of $O(m^2)$ candidate boxes with right-back pairs from $H^{\perp}$. 

If no two supports in $H^{\perp}$ are from the right-back staircase, then the top and the bottom support are both in $H^{\perp}$.
Wlog suppose the right support is in $H^{\perp}$.
By a similar argument as in the previous paragraph, we obtain $O(m^2)$ candidate boxes with exactly one support in $H^{\perp}$.

Summing up, there are $O(m^2)$ candidate boxes in case 5.

\subsection{Case 6}
\label{ss:cases-6}

\begin{figure}[t]
\centering
\includegraphics[scale=0.55]{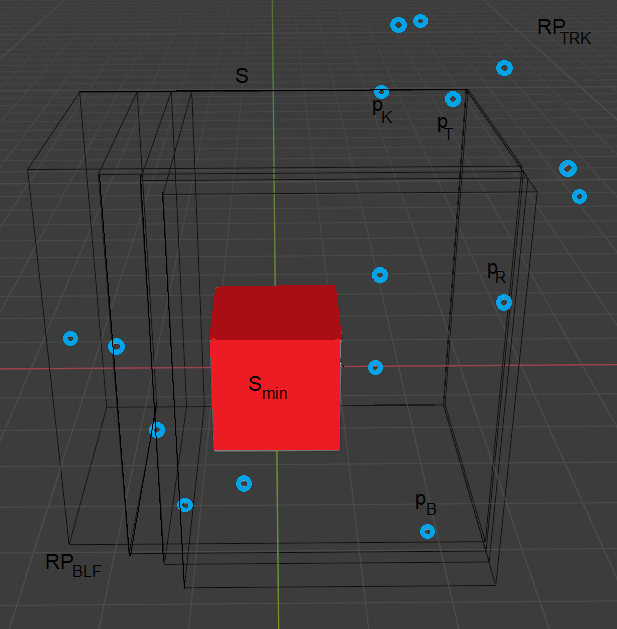}
\caption{Points $p_T \in RP_{TRK}, p_R \in RP_{TRF}, p_K \in RP_{BRK}, p_B \in RP_{BRF}$ support, respectively, the top, right, back, and bottom faces of $S$. 
There are $O(m)$ such top supports from $RP_{TRK}$, and for each of them there are $O(m)$ choices of left-front pairs $p_L, p_F$.
}
\label{fig:case6}
\end{figure}

A depiction of this case can be found in Figure \ref{fig:case6}.

Wlog suppose the 4 supports from $H$ form the top-right-back-bottom quadruple, and the top support comes from the top-right-back polyhedron.
Then the back support must come from the bottom-right-back polyhedron, the bottom support must be located in the bottom-right-front polyhedron, and the right support must be located in the top-right-front polyhedron.
For the front-left pair, we are left with $O(m)$ valid choices.
For the top support, we have $O(m)$ choices, and for every one of them, we have $O(1)$ valid bottom-right-back pairs, which give us $O(m)$ front-left pairs each.
That is, there are $O(m^2)$ candidate boxes in case 6.

\subsection{Case 7}
\label{ss:cases-7}

\begin{figure}[t]
\centering
\includegraphics[scale=0.55]{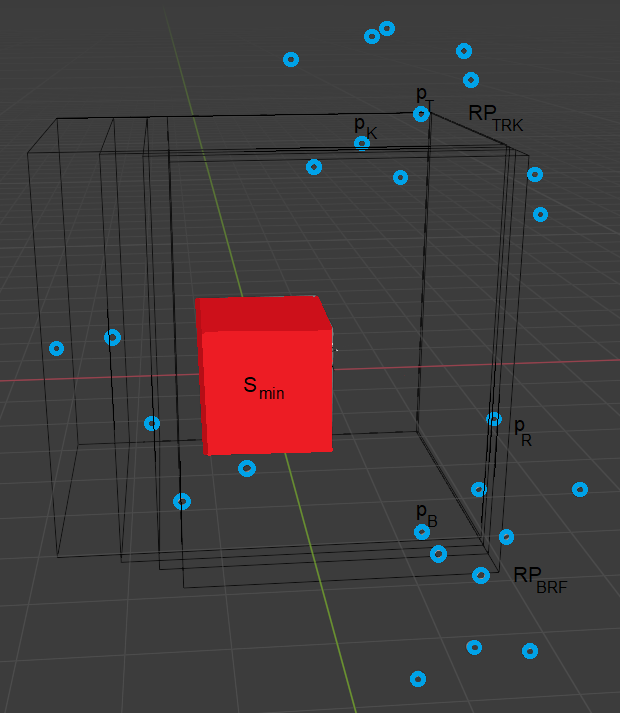}
\caption{There are $O(m)$ choices of top-back supports $p_T, p_K \in RP_{TRK}$. 
For each of them, there are $O(m)$ choices of bottom-right supports $p_B, p_R \in RP_{BRF}$.
For each such choice, there are $O(m)$ valid left-front pairs $p_L, p_F$.
}
\label{fig:case7}
\end{figure}

An illustration of this case can be found in Figure \ref{fig:case7}.

As in case 6, suppose wlog that the 4 supports from $H$ are the top-right-back-bottom quadruple.
Also let the top-back pair come from the top-right-back corner region, and the bottom-right pair come from the bottom-right-front corner region (the other case can be treated in a similar manner).
There are $O(m)$ such top-back pairs, and for each of them there are $O(m)$ such choices of bottom-right pairs.
Once top-right-back-bottom quadruple is established, there are still $O(m)$ choices of front-left pairs, as noted in case 6.
That gives us $O(m^3)$ total candidate boxes in case 7.

\subsection{Case 8}
\label{ss:cases-8}

\begin{figure}[t]
\centering
\includegraphics[scale=0.55]{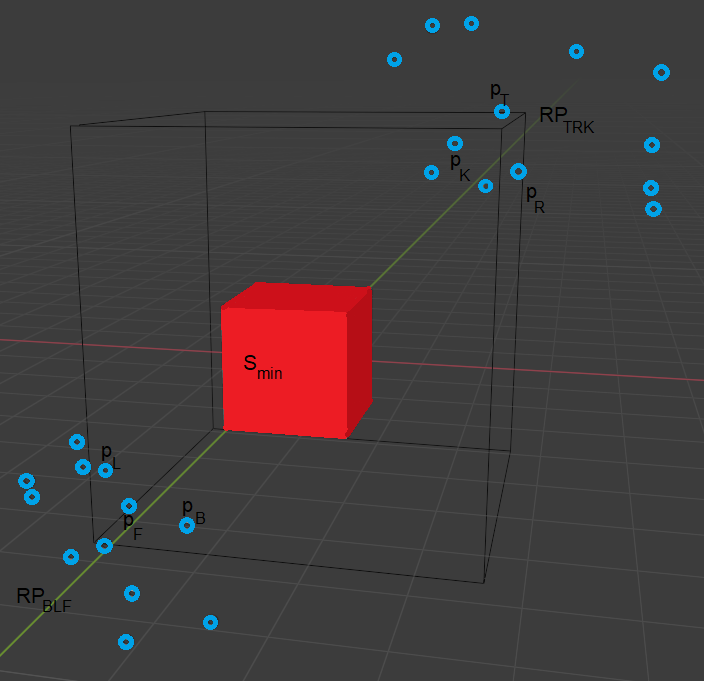}
\caption{There are $O(m)$ choices of top-right-back supports $p_T, p_R, p_K \in RP_{TRK}$. 
For each of them, there are $O(m)$ choices of bottom-left-front supports $p_B, p_L, p_F \in RP_{BLF}$.
}
\label{fig:case8}
\end{figure}

Refer to Figure \ref{fig:case8} for details.

Wlog assume the top-right-back triple is given by the top-right-back polyhedron, and the bottom-left-front triple is given by the bottom-left-front polyhedron.
Each top-right-back triples is given by an inner corner of the top-right-back polyhedron, and since the complexity of a retilinear polyhedron is $O(m)$, we get $O(m)$ potential top-right-back triples.
Similarly, we obtain $O(m)$ candidate bottom-left-front triples.
It follows that there are $O(m^2)$ candidate boxes in case 8.

\subsection{The Data Structure}
\label{ss:ds}

In order to find each candidate box in $O(1)$ time, we need the ability to find the support of a face in $O(1)$ given the supports of two adjacent faces.

\begin{figure}[t]
\centering
\includegraphics[scale=0.55]{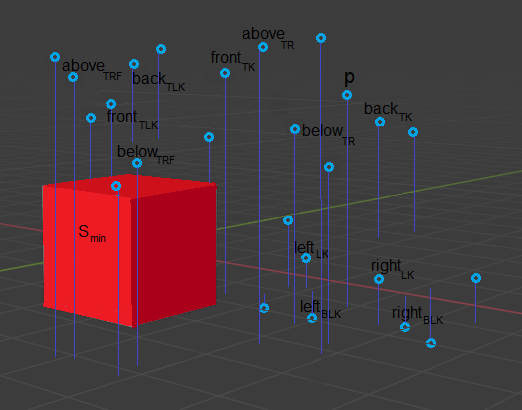}
\caption{Pointers $above, below, left, right, front, back$ for one corner region and one edge region are shown for point $p$.
To avoid convoluting the figure, we omit the argument $p$ and simply write e.g. $above_E$ instead of $above_E(p)$.
Projections of blue points onto the $XY$ plane are shown in dark blue.
}
\label{fig:ds}
\end{figure}

Thus, for each blue point $p$, we need to store the following pointers, as displayed in Figure \ref{fig:ds}.

\begin{itemize}

\item for each edge region $E$, a pointer $above_E(p)$ to the lowest point in the staircase of $E$ above $p$

\item for each corner region $C$, a pointer $above_C(p)$ to the lowest point in the polyhedron of $C$ above $p$

\item for each edge region $E$, a pointer $below_E(p)$ to the highest point in the staircase of $E$ below $p$

\item for each corner region $C$, a pointer $below_C(p)$ to the highest point in the polyhedron of $C$ below $p$

\item for each edge region $E$, a pointer $right_E(p)$ to the leftmost point in the staircase of $E$ to the right of $p$

\item for each corner region $C$, a pointer $right_C(p)$ to the leftmost point in the polyhedron of $C$ to the right of $p$

\item for each edge region $E$, a pointer $left_E(p)$ to the rightmost point in the staircase of $E$ to the left of $p$

\item for each corner region $C$, a pointer $left_C(p)$ to the rightmost point in the polyhedron of $C$ to the left of $p$

\item for each edge region $E$, a pointer $front_E(p)$ to the furthest point in the staircase of $E$ in front of $p$

\item for each corner region $C$, a pointer $front_C(p)$ to the furthest point in the polyhedron of $C$ in front of $p$

\item for each edge region $E$, a pointer $back_E(p)$ to the nearest point in the staircase of $E$ behind $p$

\item for each corner region $C$, a pointer $back_C(p)$ to the nearest point in the polyhedron of $C$ behind $p$

\end{itemize}

After computing the staircases and polyhedra, the pointers can be computed in $O(1)$ time each by traversing the staircases and polyhedra, and they are never updated.
Since we store $O(1)$ pointers for each blue point, we only need $O(m)$ additional time to compute all the pointers.
After computing all the pointers, we only require $O(1)$ time to compute each candidate box.

Since only case 7 may require $O(m^3)$ processing time, from now on, we shall focus on case 7.

Let $M$ be a tensor whose values are volumes of boxes with the top-back pair given by the row index, the bottom-right pair given by the column index, and the front-left pair given by the depth layer index.

It has been proved \cite{Smawk} that all row-maxima of a totally (inverse) monotone matrix $A$ can be computed in $O(m)$ time.
If we can prove that a depth slice of $M$ (layer) either is totally (inverse) monotone, or can be transformed into a totally (inverse) monotone matrix, then we would find all row-maxima of a layer in $O(m)$ time.
By repeating the process for each layer, we find the maxima of all rows of all layers in $O(m^2)$ time, and store them in an associative list $L$ that maps the row, collumn, and layer index in $M$ to the volume of the box. 
After that, we simply traverse $L$ and report the candidate box having the largest volume.

\begin{lemma}
For every layer $k$ of $M$, the depth slice $M[k]$ can be transformed into a totally (inverse) monotone matrix $M'[k]$.
\end{lemma}

\textbf{Proof}.
First note that, since not all bottom-right pairs may occur with a given top-back pair and the $k$-th front-left pair, some portions of $M[k]$ may be undefined.
However, the defined portion of $M[k]$ is contiguous.
If we could prove that the defined portion of $M[k]$ is totally (inverse) monotone, we would then pad each side of $M[k]$ with real values, as indicated in \cite{Armaselu-arXiv}, to produce a totally (inverse) monotone matrix $M'[k]$.
Indeed, since the front-left pair is the same for all of $M[k]$, and all of the other supports come from a single halfspace $H$, 
one can project these supports on the plane supporting $H$ to form, for each candidate box, a candidate rectangle supported by the respective top-back-bottom-right quadruple.
The largest boxes defined by $M[k]$ then correspond to the largest rectangles among these candidate rectangles.
Thus, we transform $M[k]$ by replacing volumes of boxes by areas of their corresponding rectangles.
The total monotonicity of the transformed $M[k]$ consisting of areas of rectangles follows from \cite{Mozes}.
\qed

Note that, in order to avoid quadratic computations of the entries of $M'[k]$ for each $k$, we need only evaluate $M'[i, j, k]$ whenever queried, rather than storing the whole matrix in memory \cite{Armaselu-arXiv}.
Since only a linear number of such queries are performed \cite{Armaselu-arXiv, Mozes}, it follows that all row maxima of $M'[k]$ are computed in $O(m)$ time.
Thus, we have the following result.

\begin{lemma}
All candidate boxes in case 7 can be computed in $O(m^2)$ time.
\end{lemma}

We are now in a position to state the following.

\begin{theorem}
The MBSB of a blue point set $B$ and a red point set $R$ in three dimensions can be computed in $O(m^2 + n)$ time.
\end{theorem}

\textbf{Proof}.
We compute $S_{min}$ and $S_{max}$ in $O(m + n)$ time \cite{Armaselu-CCCG}.
We then compute the staircases of all edge regions in $O(m \log m)$ time, and the polyhedra of all corner regions in $O(m \log m)$ time.
While computing the staircases, we also compute all needed pointers using an additional $O(m)$ time.
After having computed the pointers, we spend $O(1)$ to compute each candidate box in all cases except case 7.
Since there are $O(m^2)$ boxes, we spend $O(m^2)$ in all of these cases.
By Lemma 2, we also spend $O(m^2)$ time in case 7.
Thus, we spend a total of $O(m^2 + n)$ time.
\qed

\section{Conclusion and Future Work}
\label{s:conclusion}

We addressed the problem of finding the maximum volume bichromatic separating box among $n$ red points and $m$ blue points in 3D, and provided an algorithm that takes $O(m^2 + n)$ time.
We leave for future work computing the maximum volume bichromatic separating box of arbitrary orientation. 

\section*{Acknowledgement}
The author would like to thank Dr. Ovidiu Daescu for the useful discussions.
\label{s:ack}


\small
\bibliographystyle{abbrv}

\end{document}